# Motion-Compensated Handheld Quantum Key Distribution System

Hyunchae Chun[1], Iris Choi[2], Grahame Faulkner[1], Larry Clarke[3], Bryan Barber[3], Glenn George[3], Colin Capon[3], Antti Niskanen[4], Joachim Wabnig[5], Dominic O'Brien[1] and David Bitauld[5]

[1]Department of Engineering Science, University of Oxford, Oxford, OX1 3PJ, United Kingdom
[2]Department of Physics, University of Oxford, Oxford, OX1 3PU, United Kingdom
[3]Bay Photonics Ltd., Freshwater Quarry, Brixham, Devon, TQ5 8BA, United Kingdom
[4]Nokia Technologies, Karaportti 3, 02610 Espoo, Finland
[5]Nokia Technologies, Broers Building, 21 JJ Thomson Avenue, Cambridge, CB3 0FA, United Kingdom

**Mobile devices have become an inseparable part of our everyday life. They are used to transmit an ever-increasing amount of sensitive health, financial and personal information. This exposes us to the growing scale and sophistication of cyber-attacks. Quantum Key Distribution (QKD) can provide unconditional and future-proof data security but implementing it for handheld mobile devices comes with specific challenges. To establish security, secret keys of sufficient length need to be transmitted during the time of a handheld transaction (~1s) despite device misalignment, ambient light and user's inevitable hand movements. Transmitters and receivers should ideally be compact and low-cost, while avoiding security loopholes. Here we demonstrate the first QKD transmission from a handheld transmitter with a key-rate large enough to overcome finite key effects. Using dynamic beam-steering, reference-frame-independent encoding and fast indistinguishable pulse generation, we obtain a secret key rate above 30kb/s over a distance of 0.5m under ambient light conditions.**

Guaranteed by the laws of physics, Quantum Key Distribution (QKD)[1–3] provides an ultimate level of security, and uniquely it is the only technology known to-date that can monitor eavesdropping activities. Since its inception in 1984, developments of QKD has been largely focussed on securing large-scale infrastructures[4–6] using long distance fibre transmission[7–11] and free space transmission between fixed terminals[12–14] QKD transmission has also been demonstrated from large mobile vehicles such as trucks[15], planes[16] and air balloons[17]. Today, there is increasing demand for security in handheld devices such as mobile phones and wearable technologies. A QKD protected link between handheld devices and the terminal would provide the keys for encryption with verifiable security for any wireless communication system, ranging from indoor wireless networks (e.g. Wi-Fi) to access control and near field communications (NFC) mobile payment applications. This technology could also prevent ATM skimming attacks, where a $2 billion loss worldwide in 2015[18] was estimated, by transmitting the quantum encryption key from the mobile device securely to the ATM terminal. A short-range free-space implementation of QKD could address these security challenges and provide a high degree of security.

For a handheld QKD system to be practical, it needs to be compact and low-cost[19–21], and crucially, it must also be able to transmit a secure key in a time suitable for a handheld transaction (~1s). If the size of the transmitted key is too small it is not mathematically secure, so during this short span of time, a number of qubits large enough to overcome finite key effects must be exchanged[22–24]. The system must therefore provide a stable communication link taking into account the user's unavoidable hand movement, which lead to translations and rotations of the mobile device. Previous work by Mélen, G. et al.[21] obtained intermittent key transmission from a handheld device by using one-side beam-steering and mechanical waveplate rotation in the receiver. Here we addressed these issues by using a novel agile dual-MEMS mirror-based beam-steering system combined with a reference frame independent (RFI) QKD protocol to provide wider angular, translational tolerance and rotational independence respectively. This results in the first stable quantum link from a handheld device able to transmit a secure key in less than a second.

In order to compensate for hand-held operation instabilities, we characterised typical hand-movement and accordingly designed a system with optimal latency and tolerance. Based on this evaluation, we implemented the system using MEMs mirrors for beam-steering at both ends of the link. Their orientation was controlled by a tracking system using LED beacons and Position sensing detectors (PSD) at both ends of the steering system. This implementation allowed us to obtain a wide angular coverage and minimise latency, thus guaranteeing the robustness against hand movements.

The RFI QKD protocol proposed by Laing, A. et al. [25], and experimentally demonstrated by Wabnig, J. et al.[26] allows polarization encoding without the requirement of aligning the polarization bases. In this protocol the qubits are encoded and measured in three polarization bases (horizontal-vertical, diagonal-anti-diagonal, circular left-circular right). Only one of those bases needs to have a fixed alignment between the transmitter and the receiver. As the circular basis is unaffected by relative rotations of the transmitter and the receiver in polarization encoding, it is therefore used to transmit the secret keys. The two linear bases can have any relative alignment and are used to assess the security parameters of the quantum channel. Based on this protocol, we demonstrated a practical quantum secured wireless link between a terminal and a handheld device using a steerable optical link.

## Results

**QKD transmitter and receiver modules.** The RFI QKD transmitter and receiver modules are represented in Figs. 1a and 1b respectively. RFI QKD is based on the transmission and detection of qubits encoded in three bases. Here the qubits are implemented by polarization encoded faint pulses. In order to produce these 6 optical states, we use 6 resonant-cavity Light Emitting Diodes (RCLEDs, 650nm). The light produced by each of the RCLED is collimated and transmitted through a set of Polarizing Beam Splitters (PBS), non-polarizing Beam Splitters (BS) and waveplates (WP). The polarization of the light produced by each RCLED is thus purified and rotated in order to exit the transmitter in one of the 6 required polarization states. The optical signal is then attenuated by a neutral density (ND) filter to obtain faint pulses. In order to avoid side-channel attacks[27] the optical states originating from all the RCLEDs must be identical apart from their polarizations. This means that their spatial, spectral, and time profiles should be identical. Spatial and spectral indistinguishabilities are achieved by using a spatial filter with a diffraction limited pinhole and a spectral filter with a passband narrower than the sources' emission spectra (see Methods). Fig. 2a shows the resulting spectra from the six sources. Time profiles of the light emitted by each source was measured by a single photon detector collecting photons at the output of the transmitter. The resulting statistics, shown on Fig. 2b, exhibits a very good timing overlap.

In the demonstration the RCLEDs were driven by a 6-channel pattern generator. Every 4ns (i.e. a 250MHz repetition rate), a 1ns pulse is generated in one of the six channels, this channel being determined by a pseudo-random pattern. The voltage level driving each channel is adjusted to ensure that the output power was identical for the 6 polarizations. The power was then reduced to 0.07 photons per pulse with a neutral density (ND) filter. This value was chosen to obtain a sufficiently small probability for a pulse to contain more than one photon (see Methods).

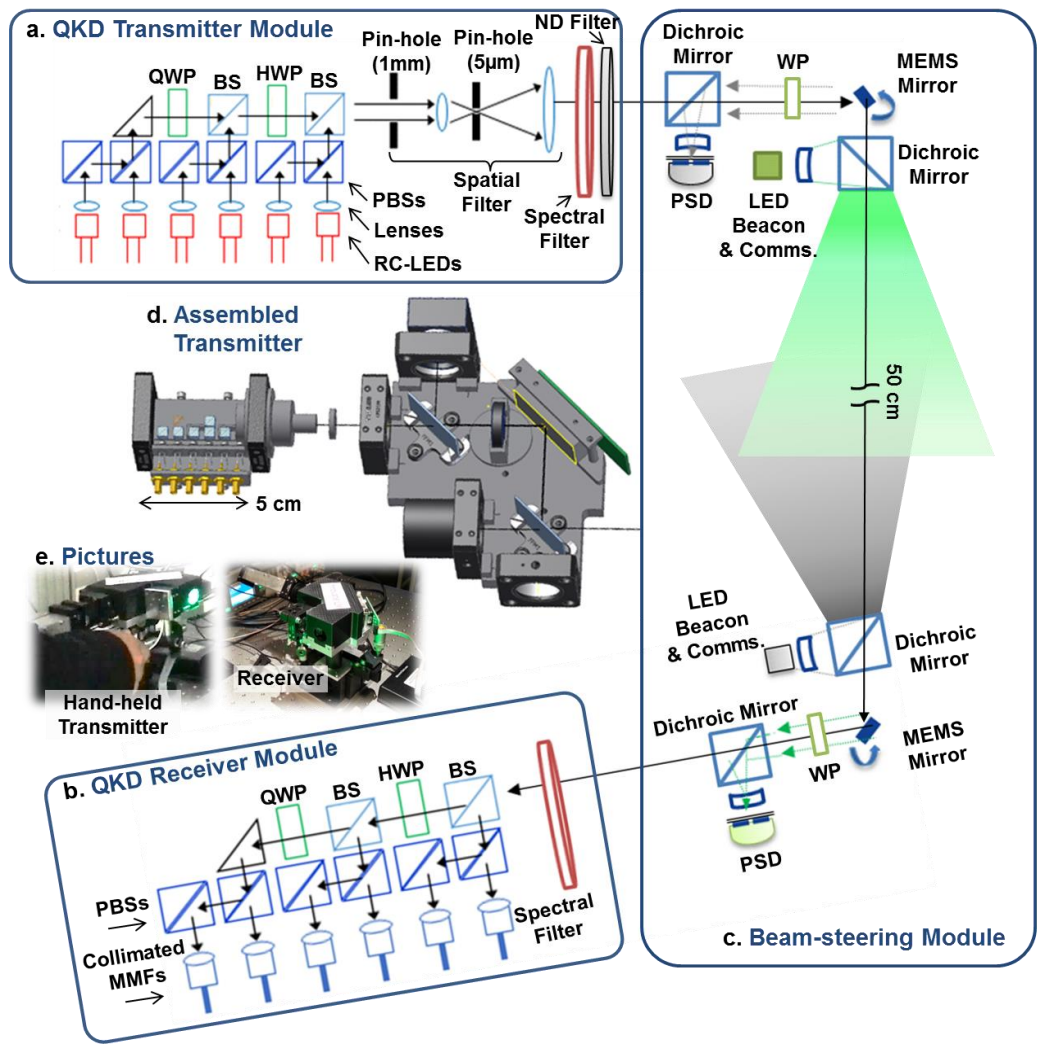

**Figure 1 | Hand-held QKD device and terminal. (a), (b)** RFI protocol based QKD transmitter and receiver module. **(c)** A schematic representation of beam-steering module and 50cm optical link. **(d)** Assembled transmitter showing the QKD and steering module. **(e)** Experimenter holding the hand-held transmitter. Also showing the receiver module in use.

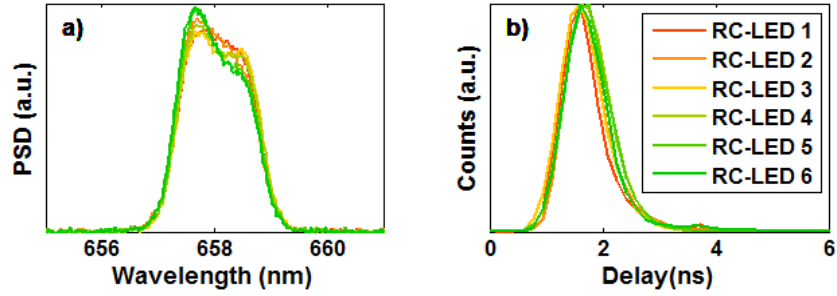

**Figure 2 | RC LED characteristics. (a)** Optical spectrum of the light generated by each RC-LED after going through the spectral filter. **(b)** Arrival time statistics for photons generated by each RC-LED. The x-axis is the delay between the synchronisation signal

The receiver module (Fig. 1b) has a similar structure to the transmitter. First, the light is spectrally filtered with the same passband as the transmitter filter. Then, light propagates through the same PBS, BS and WP arrangement as the transmitter, which splits the light according to its polarization and distributes it to six different channels. Light is then coupled into six lensed multimode fibres, which are connected to the Silicon Avalanche Photo-Detectors (SiAPD) used for photon counting.

**Agile beam-steering module.** The beam-steering module must compensate for hand movement, and ensure that the angular misalignment of the transmitter and receiver remains within the field of view of the QKD communications system. The angular fluctuations of typical hand movements were analysed by measuring the position of a hand-held laser pointer spot as a function of time. Fig. 3a shows the complementary cumulative distribution function (CCDF) of the angle created by hand-movement. This indicates the speed at which any correction must be made to the field of view of the optical system. Our receiver's the field of view, which is set by the numerical aperture of the lensed multimode fibres and associated optics, is 0.1 degree. This means that corrections must be made within approximately 42ms.

Different techniques can be used to allow the terminals to locate each other before communication. The QKD signal is too weak for tracking. Using wavelength-division-multiplexing, two beacon signals are added to system to allow simultaneous tracking and QKD operation. To minimise the system latency, a scheme using independent LED beacons is used here. Each terminal has a beacon that indicates its position, and a position sensitive detector (PSD) to measure the location of the beacon on the other terminal. This dual beacon system allows the two terminals to track each other independently, resulting in low-latency. Fig. 1c shows the system we used for dynamic beam-steering. A green LED beacon (520nm) aligned with the QKD optical mode is used in the transmitter, to indicate its position to the receiver, allowing it to perform adjustment of its field of view. For the receiver to evaluate the transmitter's position, light from the beacon enters the receiver and is diverted to a PSD via a MEMs steering mirror. The PSD then measures the angle of arrival of the beacon light, creating a signal that steers the MEMs mirror until the beacon light is centred on the PSD. This sets the receiver mirror angles so that photons coming from the direction of the QKD transmitter will be coupled into the lensed fibres. The receiver also has a co-aligned infrared LED beacon (850nm), which allows the transmitter to orientate its MEMs mirror and steer the QKD photons towards the receiver. In operation, coverage of +/-4° was achieved, as shown in Fig. 3b.

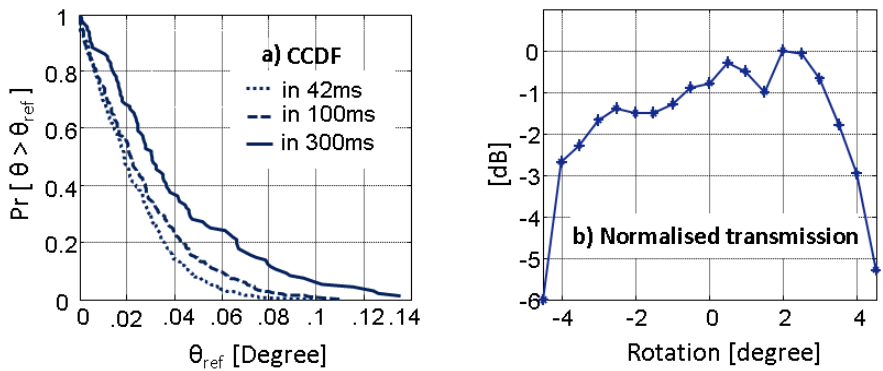

**Figure 3 | Hand-movement statistics and beam-steering coverage. (a)** Complementary cumulative distribution function (CCDF) of the maximum angle deviation created by hand-movement in different time durations of 42ms, 100ms, and 300ms,.where $\theta_{ref}$ is reference angle used for the CCDF **(b)** Normalised transmission loss by beam-steering with respect to the rotation of transmitter angle.

**QKD performance evaluation.** Initial static tests were undertaken, for different steering angles. It was found that the optical elements in the beam-steering system (and to a small extent the beam splitters inside the transmitter and the receiver) introduced an undesired birefringence leading to polarization rotations of the QKD beams. RFI QKD is tolerant to polarization rotations, but not to ellipticity alterations, so a WP was added inside the beam-steering modules to compensate for this. The induced birefringence was also dependent on the MEMS mirror angle, which varies, and therefore cannot be compensated with a static WP, but it was found that the increase in the error rate remained within 3% over the whole steering range.

In order to ensure absolute secrecy of the transmitted key, the raw key that is transmitted with the circularly polarized photons needs to be processed using error correction and privacy amplification codes. These reduce the key size to a fraction of its original size, called the secure key fraction. This secure key fraction is calculated based on correlations between the number of photons transmitted in each polarization state and the number of detections in each receiver channel, using the method described in the Methods section (based on Wabnig, J. *et al.*[26]). These calculations take into account potential polarization impurity and non-orthogonality and it includes finite key analysis.

In order to demonstrate the robustness of our system against axial rotations (i.e. rotations around the transmission direction) we introduced a rotating half-wave plate between the emitter and the receiver, which is equivalent to a rotation of the transmitter by an angle equal to twice the value of the waveplate angle. The secure key fraction as a function of the equivalent rotation angle is shown on Fig. 4. Here the secure key fraction was calculated for 0.5s key transmissions. We can see that not only transmission of secure keys was possible for any angle but the secure key fraction remained between 40% and 60% over a 360 degree rotation. The secure key fraction was also calculated by using the BB84 protocol with the same photon counts, and therefore keeping two bases out of three to calculate the key fraction. In this case we can see that the system performances would drop very quickly with the transmitter angle. RFI protocols is superior to BB84 for each point except for the angle of 170 degrees. This is due to the fact that the calculation we used for the BB84 curves assumes perfect polarization orthogonality, while our RFI calculation takes the most pessimistic possibility without assuming perfect polarization.

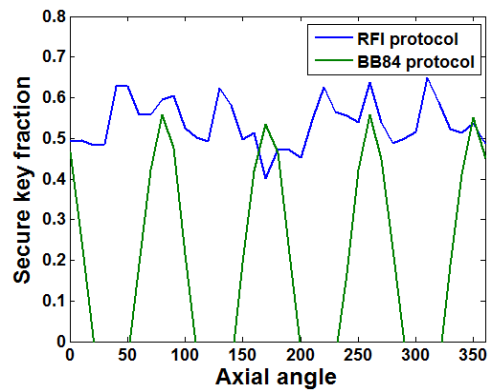

**Figure 4 | Robustness of RFI protocol against axial rotation.** The secure key fraction (ratio between the raw key length and the error corrected, privacy amplified key length) is plotted as a function of the relative axial angle between the transmitter and the receiver. The values were calculated from a 0.5s data transmission for each point, for a fixed position of the terminals. The relative axial angle between the transmitter and the receiver was simulated by rotating a half-wave plate between the terminals.

Rotations around the other axes induced small modifications of the photons' polarization but with minor effects on the secure key fraction. The secure key fraction over the whole possible range of deflection of the mirrors remained between 35% and 65%.

The system operates in normal ambient light conditions (>400 lux from incandescent light bulbs). The detector dark count rate is 370 counts/s per detector, and the noise due to ambient light was 600 photons/s per detector. The beacon LEDs additionally contribute approximately 570 photons/s per detector. These numbers are small compared to the ~1M photons/s that are detected for the QKD link, and induced error rate due to these impairments is negligible.

Quantum Key Distribution from a handheld device was demonstrated by holding the transmitter and pointing it at a static receiver. The whole optical assembly including the QKD and beam steering modules were part of the mobile handheld device, while the electrical instruments were static and connected to the handheld part by cables. Two sets of experiments were performed. First, in order to illustrate the fluctuations of the asymptotic key rate in real time, measurements were performed while the experimenter holding the device performed a series of voluntary movements. To calculate this asymptotic key rate, short (4ms) samples of QKD transmission were collected every 100ms, which is the time required to process the photon arrival times. Fig. 5a shows the vertical and horizontal angles of the two MEMS mirrors as well as the axial angle. Mirror angles were derived from the applied voltage while the axial angle was derived from the photon statistics, which depend on the polarization angle (see Methods). The alignment system was switched on after 4s, and no data was received before this point. Then we can see that the experimenter moves the transmitter up, down, left, right and finally performs an axial rotation.

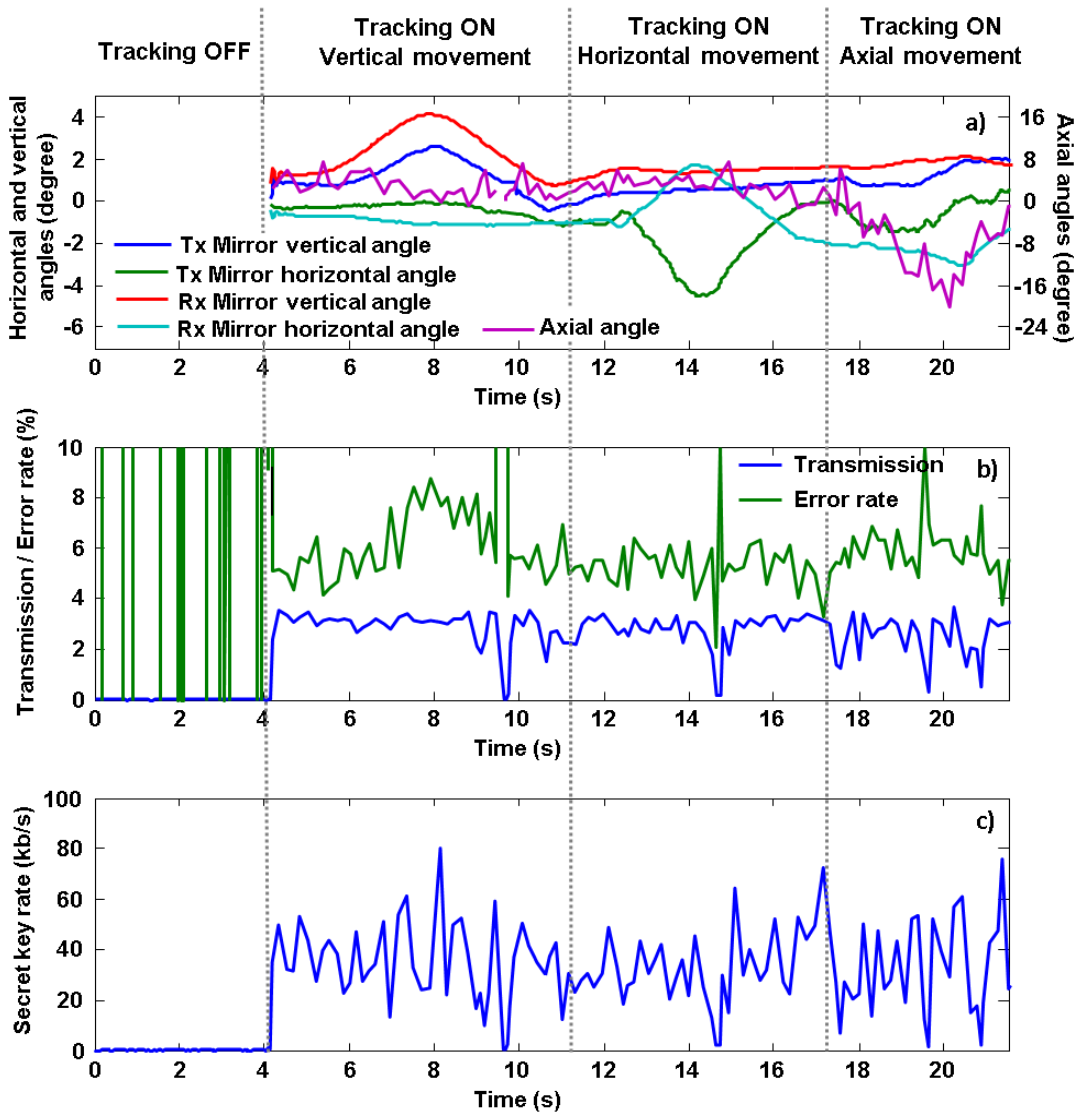

**Figure 5 | Hand-held QKD performance.** These measurements were performed while the experimenter holding the QKD transmitter displaces it in the directions shown. The motion compensation system was switched on after 4s. **(a)** Vertical and horizontal angles of the two MEMS mirrors as well as the axial angle between the two terminals. The mirror angles were retrieved from the applied voltage, while the axial angle was calculated from the QKD data. **(b)** Photon transmission and error rate. The transmission is defined as the number of detected photons divided by the number of emitted photons (i.e. 0.07 x 250MHz). The error rate is calculated for photons that were both transmitted and detected in the Z basis. **(c)** Estimation of the asymptotic key rate as a function of time. Each point was calculated from a 4ms transmission sample.

Fig. 5b shows the transmission and error rate as a function of time. The transmission is defined as the total number of detected photons divided by the total number of emitted photons (i.e. 0.07 x 250MHz), while the error rate only takes into account the photons used for the secure key (i.e. photons transmitted and received in the Z basis). After starting the steering system, the transmission stabilizes at around 3% with a few short drops and the error rate is approximately 6%. Fig. 5c shows an estimation of the asymptotic key rate as a function of time, indicating a rate of approximately 30kb/s with very few drops to zero. Fluctuations in this rate are due to the short measurement samples.

Second, in order to demonstrate the performance of our system during a real handheld transaction, including the finite key effects, a series of 14 separate measurements was performed with a duration of 0.5s each, while the experimenter was holding the transmitter without any intentional movement. The average value of the secure key rate, after the finite key effect is taken into account, was 45kb/s with a standard deviation of 10kb/s. These results show that our system can reliably transmit secret keys long enough to overcome the finite key limitations in a very short time.

**Discussion**

In this paper, we have successfully constructed the first quantum wireless prototype for handheld usage fulfilling finite key requirements. The beam-steering technique used maintained a motion-stabilized communication link, which allowed sufficient secure keys to be transmitted with a high throughput under hand-movement and typical ambient light levels.

A number of improvements are possible. An increased key rate could be achieved by using a decoy state protocol[28,29], For full mobile device integration, further miniaturization of the emitter could be obtained by fabricating the RCLEDs on a single chip and combining the polarizations with photonic waveguides as previously demonstrated with VCSELs[20]. The steering system could also be miniaturized by taking advantage of the fast improvements and size reductions of laser-based pico-projection systems[30].

The success of wireless QKD technology could introduce an unprecedented level of data security to merchants and customers alike. Making QKD economical and practical enough is essential to overcome the adoption barrier before security threats start to appear with the increase of computing power and efficiency of hacking techniques. This demonstration of handheld quantum wireless prototype represents a major step towards real-world quantum security for mobile applications.

**Methods**

**Fabrication of QKD transmitter and receiver modules.** The light sources used in the receiver were resonant cavity LEDs (RCLEDs) with a central wavelength of approximately 656 nm and a spectral width of 7nm. In order to ensure identical transmission spectra, a spectral filter with a passband (1.5 nm centred at 658 nm) narrower than the RCLED emission spectrum is used to select the light that will be transmitted. The choice of RCLEDs over laser diodes or LEDs is important for this application. First, RCLEDs have a wider spectrum than lasers, without multiple mode structure. The disparities between the central wavelengths of the different RC-LEDs being smaller than the spectral width, the light going through a narrow band filter will be similar for each RC-LED (Fig. 2a). This is very difficult to achieve with lasers because they need to be tuned with a tolerance smaller than their individual modes' width and filtered with a narrower filter. Second, RCLEDs can be modulated faster than LEDs, making it possible to transmit a larger number of photons compared to the constant detector noise rate and thus, increasing the key rate. The spatial filter was composed of a first aperture with a 1mm diameter followed by a focusing lens (f=6.24mm) and a second aperture with a 5 micron diameter (comparable to the Airy disk diameter). This ensures any difference in mode profile or propagation direction between sources is eliminated. The light is then collimated with an f=11mm lens. Both the transmitter and receiver base plates were fabricated using aluminium terminated with modified 16mm cage plates to house the spatial filter and to enable easy integration with the beam-steering system. Alignment of the receiver's lensed multimode fibres was achieved using spherical bearings to give the degrees of freedom required. All components were fixed in place using UV cured adhesives. A 10dB neutral density (ND) filter was added after the transmitter to obtain the required faint pulse intensity. Waveplates (WP) were added in the transmitter and receiver steering modules to correct for their induced birefringences. The WP angle around vertical axis as well as the relative angle of the QKD and steering modules were adjusted to obtain perfectly circular polarizations in the Z basis. We also took advantage of the fact that the additional WP acted essentially as a quarter waveplate to swap the X and Z basis by rotating the QKD receiver module by 45 degree. This allows us to have the Z basis (used for the secure key) nearer to the receiver input. By construction this made the detection probability and the polarization purity higher than when the Z basis is at the opposite end of the receiver.

**Beam-steering module**. Commercial Position sensing detectors (PSDs) and MEMS mirrors (ϕ=4.2mm) were used. For the mirror, larger area leads to longer latency, mainly due to the inertia, but leads to a better reliability by reflecting more off-axis beams. This PSD and MEMS mirror combination enabled a low latency suitable for an agile correction of hand-movement. Diverging beacon LEDs (LED Engine Inc. 520nm for transmitter and 850nm for receiver) were used. The wavelengths were chosen to ensure enough separations between the weak QKD signal at 658nm and the strong beacon signals, and to help aiming to the receiver with 520nm visible light. Dichroic beam splitters of FF750 and FF562 (Semrock Inc.) were used.

**QKD secure key rate calculations.** Our security analysis takes into account that, in real implementations, not only reference frames of the transmitter and the receiver can be rotated with respect to each other, but also that there is always a degree of misalignment within a reference frame. This induces non-orthogonality within a basis and mutual bias between bases. Our analysis also takes into account differing detector efficiencies. This is achieved by using an explicit device model and minimising the key rate over possible model parameters. The calculation of the secret key fraction can be posed as an inference problem: given the measurements and a device model, what is the maximum amount of information an eavesdropper can possess about the distributed key?

Following the derivation in Wabnig, J. *et al.*[26] we can express the secret key fraction as

$$r = S_{min} - h\left(\frac{1 - C_{ZZ} + \sigma\delta C_{ZZ}}{2}\right).$$

We take into account imperfect measurement devices that can lead to a large number of additional parameters, such as non-orthogonalities in the preparation and measurement bases and detector efficiencies, which can be collected into the vector $\boldsymbol{\alpha}$. The usable entropy is obtained as the minimum over all parameters $\boldsymbol{\alpha}, \lambda_{1,2}$

$$S_{min} = \min_{\alpha,\lambda_1,\lambda_2} S_U(\lambda_1, \lambda_2).$$

The parameters have to obey the constraints imposed by the observations

$$f_i(\boldsymbol{m}) - \sigma\delta f_i(\boldsymbol{m}) \leq f_i[\boldsymbol{q}(\boldsymbol{\alpha}, \lambda_1, \lambda_2)] \leq f_i(\boldsymbol{m}) - \sigma\delta f_i(\boldsymbol{m}),$$

where $\boldsymbol{m}$ is a matrix containing all relevant detector counts, $\boldsymbol{q}$ is the corresponding probability of observing a detector count according to the device model, the $f_i$ are the functions defining the different constraints, the $\delta f_i$ are their corresponding variances and $\sigma$ is chosen to give a certain probability that the estimated usable entropy is too high. In our parameter estimation step we use a set of 21 constraints consisting of 9 correlation functions $C_{AB}$, $A, B = X, Y, Z$, the six probabilities that a photon was prepared in a certain polarisation direction $P_{A\pm}$, $A = X, Y, Z$ and the six probabilities to detect in a certain detector $D_{B\pm}$, $B = X, Y, Z$. For each function $f_i$ we can give the standard deviation $f_i$. To obtain the secret key fraction we need to subtract the observed relative entropy in the key bases from the usable entropy

The full set of measurements enables us to calculate a reference frame independent key rate. From the detector counts we can construct 9 correlation functions

$$C_{AB} = \frac{m^{AB}_{++} + m^{AB}_{--} - m^{AB}_{+-} - m^{AB}_{-+}}{m^{AB}_{++} + m^{AB}_{--} + m^{AB}_{+-} + m^{AB}_{-+}}, \quad A, B = X, Y, Z,$$

where the $m^{AB}_{\pm\pm}$ are the four different detector counts given that the qubit was prepared in direction $A \pm$ and detected in direction $B \pm$. In the case of orthonormal preparation and measurement bases, these can be directly identified with the qubit correlation functions.

Each preparation and each detector are associated with a direction on the Poincaré sphere, given by a unit vector. Since we aim to use the z-basis for the key bits we can identify two preparation directions with the $\pm z$ direction on the Poincaré sphere, without overestimating the secret key fraction. Each direction is parametrised by two variables (e.g. azimuth and polar angle), resulting in a total of 20 free parameters. Different absorption may occur in the preparation channels, and similarly detectors may have different efficiencies. With six possible preparation directions and six possible detection direction this adds another set of 12 parameters. The quantum channel can be represented by a two parameter two qubit density matrix in a simplification over the more commonly employed Bell diagonal density matrix. This results in a model with 34 free parameters. The secret key fraction is then obtained as the minimum over all parameters that fulfil the constraints imposed by the measurements, e.g. from the correlators $C_{AB}$; in our case a set of 21 constraints. For the number of detector counts approaching infinity the constraints are equalities, but for a finite number of observations the value can lie within an interval determined by the number of counts and the desired uncertainty. A small number of counts will lead to larger uncertainty in the correlation function and therefore to lower results for the secret key fraction. Finally, since we don't use single photon sources in the transmitter, we need to consider the occasional transmission of multi-photon pulses, which could be exploited by an adversary able to non-destructively detect those pulses and extract one of the photons. According to Poisson statistics, the probability for a pulse to contain 1 photon is approximately 0.07 and the probability to have more than one photon is approximately $0.07^2/2$. This means that amongst the detection events we use to build the secret key, a proportion equal to 0.035 is accessible by the adversary, which is very small compared to the overall secure key fraction (~0.5). In any case, privacy amplification and error correction will use a key fraction slightly lower than the theoretical limit to ensure an extra guaranty of secrecy.

**Axial angle estimation.** The correlators defined in the above section can also be used to estimate the relative axial angle between the QKD bases of the transmitter and the receiver. Several combinations of correlators can be used to calculate the angle. This calculations based on single photon counts are intrinsically noisy. In order to reduce this noise a median of the different possible ways to retrieve the angle Ω was calculated:

Ω= median ( atan2($C_{XY}$,$C_{XX}$)+45, atan2($C_{YX}$,$C_{XX}$)+ 45, -atan2($C_{YX}$,$C_{YY}$) -135, -atan2($C_{XY}$,$C_{YY}$) -135).

The 45 degree rotation is due to the rotation of the QKD module in the receiver as explained in the "Fabrication of QKD transmitter and receiver modules" section of the Methods. This calculation being based on the photon counts, a low transmission doesn't allow us to retrieve an accurate angle. This is why there is a missing section in Fig. 5a around 9.5 seconds. This corresponds to a transmission drop.